# The Thing with E.coli: Highlighting Opportunities and Challenges of Integrating Bacteria in IoT and HCI

**Raphael Kim**
Queen Mary University
London, United Kingdom
r.s.kim@qmul.ac.uk

**Stefan Poslad**
Queen Mary University
London, United Kingdom
stefan.poslad@qmul.ac.uk

## ABSTRACT

With advances in nano- and biotechnology, bacteria are receiving increasing attention in scientific research as a potential substrate for Internet of Bio-Nano Things (IoBNT), which involve networking and communication through nanoscale and biological entities. Harnessing the special features of bacteria, including an ability to become autonomous - helped by an embedded, natural propeller motor - the microbes show promising array of application in healthcare and environmental health. In this paper, we briefly outline significant features of bacteria that allow analogies between them and traditional computerized IoT device to be made. We argue that such comparisons are critical in terms of helping researchers to explore human-bacteria interaction in the context of IoT and HCI. Furthermore, we highlight the current lack of tangible infrastructure for researchers in IoT and HCI to access and experiment with bacteria. As a potential solution, we propose to utilize the DIY biology movement and gamification techniques to leverage user engagement and introduction to bacteria.

## 1 INTRODUCTION

The Internet of Bio-Nano Things (IoBNT) [2] encompasses communication and network architecture that are made up of biological entities and nanoscale devices. Facilitated by the recent advancements in synthetic biology and nanotechnology, IoBNT promises a wide range of potential applications. One of the bio-nano things that are under research are bacteria. In the area of environmental sustainability, bacteria could be programmed and deployed in different surroundings, such as the sea [15] and 'smart cities' [6], to sense for toxins and pollutants, gather data, and undertake the bioremediation processes.

Similarly, in medicine and healthcare, bacteria could be programmed to treat diseases. Harbouring DNA that encode useful hormones for instance, the bacteria can swim to a chosen destination within the human body, produce and release the hormones when triggered by the microbe's internal sensor [2].

As these potential applications aim to utilize the sensing, actuating, communicating, and (biological) processing abilities of bacteria, the microbes share similarities with components of typical computer IoT devices (fig. 1). This presents a strong argument for bacteria to be considered as a living form of Internet of Things (IoT) device. And consequently, in the field of Human-Computer Interaction (HCI) and Bio-HCI [11], such shift of perception of bacteria brings a rich area for exploration.

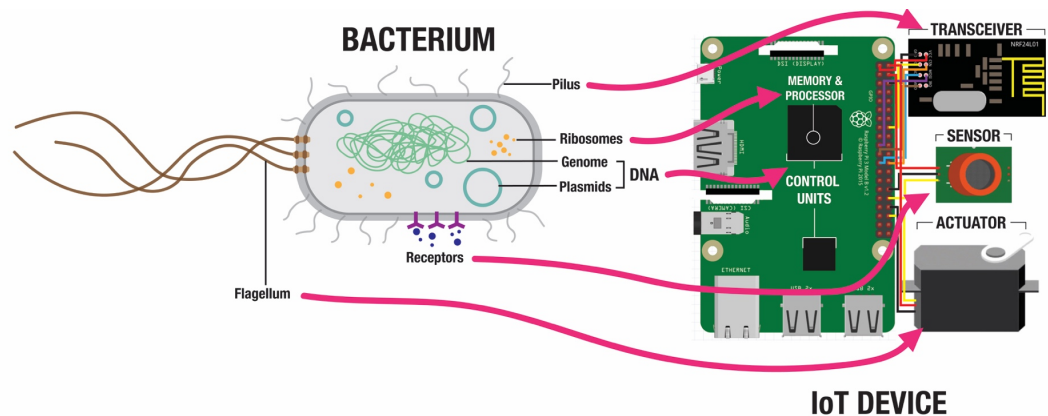

**Figure 1: Comparison between *E.coli* bacterium and Raspberry Pi-controlled IoT device. Components of the bacteria anatomy that function similarly to the computerized device are linked by pink arrows.**

## 2 BACTERIA AS IOT DEVICE

Perhaps the easiest way to explore the role of bacteria in IoT and HCI in a meaningful way, is to compare the microbes with existing computerized IoT devices in the field (fig. 1). This way, the features of bacteria can be contextualized against standardized digital models, and allow better appreciation of the species' attributes and drawbacks.

In Akyildiz *et al.* [2], an illustration depicting the shared elements of generic biological cell and a typical IoT device was presented. Here, we narrow the model further. Fig. 1 illustrates an *E.coli* bacterium (hereafter called 'bacteria'), and Raspberry Pi with add-on components, to represent the biological and digital worlds. They were deliberately chosen for this paper, as they are the common, standardized 'workhorses' of DIY biology and open source computing, respectively.

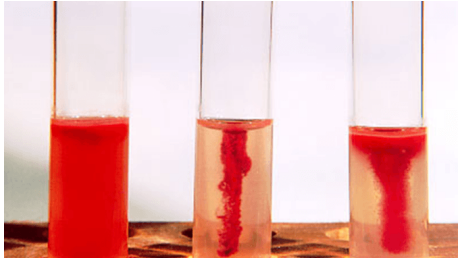

**Figure 2: Experiment outcomes showing motility of *E.coli* bacteria (in red). Various swimming patterns have been produced, depending on varying degrees of motility.**

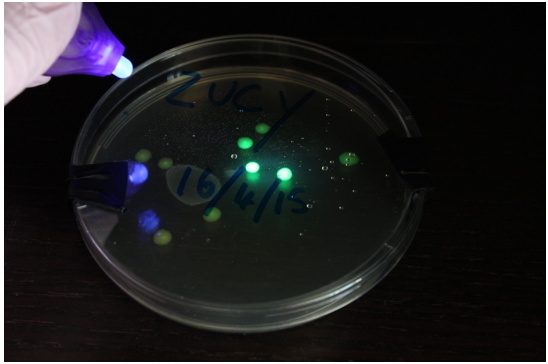

**Figure 3: A plate of *E.coli* bacteria with GFP expression, allowing them to glow green under UV light. This is an actuation of the bacteria following sensing of antibiotic stimulus.**

And as such, they both make an ideal candidate to be a cost-effective and accessible tool for meaningful engagement and learning.

Overall, whilst we acknowledge the over-simplification of both the digital and biological worlds in fig. 1, the illustration is designed to raise important questions on how bacteria could be deployed in the context of HCI investigations. These questions may arise from not only the practical differences in stability, responsiveness and even safety of interaction, but also future ethical implications of 'Internet of Bacteria Things (IoBT)' may have on our society.

## 2.1 Sensors and Actuators

Bacteria can sense a wide range of stimuli, such as light, chemicals, mechanical stress, electromagnetic fields, and temperature, to name a few. As a response to these stimuli, bacteria may interact through movement using their flagella (fig. 2), or through the production of proteins, including colored pigments such as green fluorescent protein (GFP) (fig. 3).

Given the molecular nature in which bacterial sensors and actuators operate, there can be major differences in their responsiveness, sensitivity, and stability, compared to the digital counterparts.

## 2.2 Control Unit, Memory and Processor

The DNA inside the bacteria offer storage of data and encodes instructions that can be translated into viable functions. And as such, it offers similar roles as a computer's control unit (e.g. as an entity managing 'data units' and software conditional expression units), memory (e.g. storage of embedded system data), and the processing unit (e.g. executing software instructions).

There are two main forms of DNA present in the bacteria, first as a genomic DNA, containing most instructions for cell functioning, and second as smaller circular units called plasmids. In synthetic biology, plasmids are often used to introduce a wide range of genes into the organism, making it a versatile tool for function customization and storage of new data.

## 2.3 Transceiver

As a component designed to allow both the transmission and reception of communication, cellular membrane of bacteria can be considered as a transceiver. It is involved in release and import of molecules as part of cell signaling pathways. Additionally, the bacterial pilus (fig. 1) is used for conjugation process between two cells which results in DNA exchange. Overall, such types of communication are referred as molecular communication, which forms the basis of bacterial nanonetworks.

## 3 BACTERIAL NANONETWORKS

Bacterial nanonetworks is an example of molecular communication, which has been gaining increasing attention in the IoT community [1,2,5]. Bacterial nanonetworks involve communication between bacterial communities through molecular signaling [19], and as discovered recently, through electrical fields [13]. Another method for communication is through physical movement: This involves harnessing the motile ability of bacteria, such as *E.coli*, to act as information carriers.

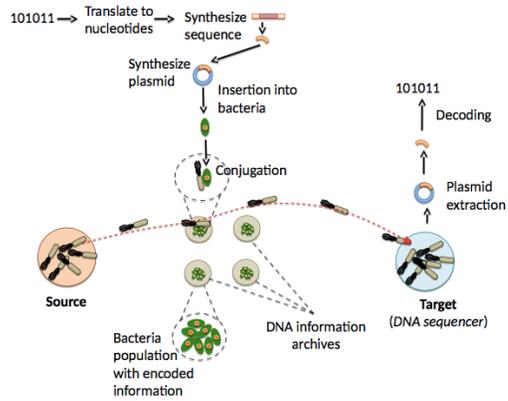

Figure 4: A schematic for bacterial communication, proposed by Tavella et al. [16]. Digital information is encoded as DNA, inserted into bacteria, which are transferred from one cell to the other, and eventually transported through swimming *E.coli*.

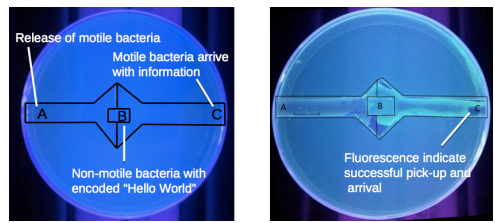

Figure 5: Experiment demonstrating successful bacterial nanonetwork and communication, by Tavella *et al.* [16]. Encoded "Hello World" message in DNA were successfully transported and the outcome is signaled by GFP expression in *E.coli*.

In such cases, the digital information is translated into DNA which can be transformed into bacterial cells, and later decoded back to the digital format. An example of its implementation, which involve conjugation between non-motile and motile bacteria, to store and transport DNA data, were demonstrated by Tavella *et al.* [16] (fig. 4, 5).

## 4 CHALLENGES AND POSSIBLE SOLUTIONS

Despite the rich framework that bacteria bring in HCI investigations, working directly with the micro-organisms may bring practical and ethical challenges. Given its biological nature, handling and manipulation requires careful consideration, which do not necessarily need to be addressed when working with computers and electronics. These include potential professional supervision, license to handle some microbes safely and legally, cost implications, and the bioethical and biosafety challenges that may arise.

Furthermore, HCI investigators may not be familiar with bacteria, and have no previous working knowledge in the field, which may pose challenges in meaningful engagement. This will be compounded by some of the fundamental concepts of bacterial physiology which can appear abstract to those without prior knowledge. In order to overcome these challenges, we propose to leverage potential HCI investigations on the DIY biology movement and gamification to aid the process. Below we outline them in further detail and explain our rationale.

### 4.1 DIY Biology

DIY Biology movement promotes increasing accessibility and affordability of tools, data, and materials of biotechnology [9]. It capitalizes on the changing economic landscape of modern biotechnology industry, represented by the continually declining cost of DNA synthesis and sequencing. Currently, tools and techniques to run small-scale experiments with micro-organisms are widely available to the general public, through various channels, including maker spaces.

Furthermore, there are supportive kits too. Educational products such as *Amino Labs* [3] cater specifically for those interested in manipulating and genetically engineering *E.coli* species. It allows creation of customized colors from bacteria, through building genetic circuits that can be triggered from a range of environmental stimuli (fig. 6,7).

Bacteria, especially *E.coli*, make an ideal tool for biohacking projects, given that they are easy to acquire, culture, and maintain. The industry standard strain *K-12 E.coli*, for example, which are used in *Amino lab*, are relatively safe to handle. They have been engineered to be non-pathogenic and difficult to spread outside of the laboratory environment. Unlike most species of bacteria, the *K-12* strain can be purchased relatively easily in the U.S and most parts of Europe.

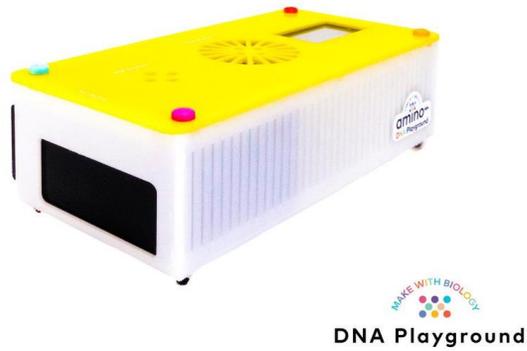

Figure 6: A DIY biology toolkit *DNA Playground* by *Amino Labs* [3]. It facilitates small-scale, simple genetic engineering experiments of bacteria.

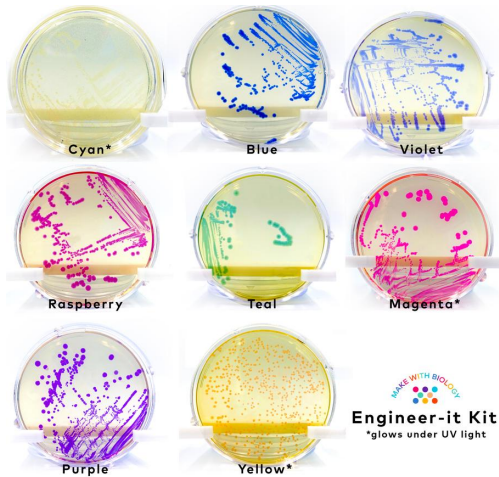

Figure 7: A range of colored proteins produced by genetically-engineered *E.coli*, made with *Engineer-It Kit* by *Amino Labs* [3].

### 4.2 Gamification of Bacteria

In the context of IoT, gamification have shown several benefits. For example, it has shown to increase engagement of new IoT applications [4], and positive shifts in human behavior (eg. travel behavior as part of a Smart City initiative [12]). Similarly, Wood *et al.*'s *GPS Tarot* is a playful, artistic tool that allows participants to learn and become aware of 'hidden technologies' in the form of embodied Global Navigation Satellite Systems (GNSS) satellites [18,19].

In a similar vein, we hypothesize that gamification of *E.coli* can aid in engagement, learning, and attitude shifts in their integration in HCI and IoT investigations. Micro-organisms have been gamified before, as a form of a biotic game, which is a hybrid bio-digital game that integrate real microbes into computer gaming platforms [14]. Overall, gamification of microbes has proven successful in terms of engagement, playing experience, and learning [7,8,10].

## 5 ETHICAL DISCOURSE

As with any potential IoT applications, ethical considerations and privacy issues in relation to user data would also apply to bacteria driven IoT systems. Interestingly, however, due to the biological nature in which such systems can operate, additional layers of ethical challenge is presented. Firstly, one of the challenges stem from the autonomous nature in which bacteria can function. As they can evolve and behave autonomously, they could pose threat to natural ecosystems, and even become pathogenic. This may not apply to educationally-deployed *K12 E.coli* strains of course, but nevertheless the possibility is worth considering from a wider perspective.

Secondly, bacterial nanonetworks rely on transfer of data (encoded in DNA) through a natural process of conjugation and cell motility. Although highly-engineered bacteria may provide efficient communication systems, ultimately, they are biological entities, which can produce unexpected outcomes (e.g. mutations). All in all, whilst the use of bacteria in IoT and HCI offer exciting opportunities, they also present fresh ethical challenges, which provide extra paths for discourse.

## 6 CONCLUSION

In this paper, we have highlighted features of *E.coli* bacteria that could enable rich exploration of microbes in the context of IoT and HCI. This was initiated by comparison of the organism with a conventional computer-driven IoT device. Furthermore, we have highlighted the current lack of tangible infrastructure for researchers in IoT and HCI to access and experiment with bacteria. As a potential solution, we have proposed to utilize the DIY biology movement and gamification techniques to leverage user engagement and introduction to bacteria. And finally, we have outlined the ethical challenges that bacteria-driven IoT systems may pose, which are brought on by the autonomous and biological (and therefore unpredictable) nature of bacteria. Such challenges offer a rich area for discussion on the wider implication of bacteria driven IoT systems.

## ACKNOWLEDGMENTS

This research was supported by EPSRC and AHRC Centre for Doctoral Training in Media and Arts Technology (EP/L01632X/1).

Image credits:
Fig. 1: Raphael Kim
Fig. 2: Tankeshwar Acharya
Fig. 3: Raphael Kim
Fig. 4: Federico Tavella
Fig. 5: Federico Tavella
Fig. 6: Amino Labs https://amino.bio/
Fig. 7: Amino Labs https://amino.bio/